\documentclass[12pt,showkeys,nofootinbib]{revtex4-2}

\usepackage{setspace}
\usepackage[utf8]{inputenc} 
\usepackage[T1]{fontenc} 
\usepackage[english]{babel} 
\usepackage{xcolor} 
\usepackage{amsmath,amsfonts,amssymb,amsthm,braket,dsfont,comment,leftindex,bbm} 
\usepackage{graphicx} 
\usepackage{subcaption,ragged2e} 
\usepackage{tikz} 
\usepackage{multirow,afterpage,tabularray} 
\usepackage{fancyhdr,orcidlink} 
\usepackage[capitalise]{cleveref} 
\usepackage{enumitem} 
\usepackage{titlesec} 
\usepackage{lipsum}
\usepackage{listings} 
\usepackage{cancel}

\newcommand{\diff}{\ensuremath{\mathrm{d}}}
\newcommand{\bal}{\begin{aligned}}
\newcommand{\eal}{\end{aligned}}

\definecolor{ao(english)}{rgb}{0.0, 0.5, 0.0}

\renewcommand{\bm}[1]{\boldsymbol{#1}}

\titleformat{\subsubsection}
{\normalfont\fontsize{9pt}{11pt}\selectfont\bfseries}
{\thesubsubsection.}
{1em}
{\centering}
\titleformat*{\subsubsection}{\normalsize\itshape}

\titleformat{\paragraph}
{\normalfont\normalsize\itshape}
{\theparagraph.}{1em}{}
\titlespacing*{\paragraph}
{0pt}{3.25ex plus 1ex minus .2ex}{1.5ex}

\begin{document}

\title{Glueballs: hadrons without quarks}

\author{\surname{Cyrille} Chevalier \orcidlink{0000-0002-4509-4309}}
\email[E-mail: ]{cyrille.chevalier@umons.ac.be}
\affiliation{Service de Physique Nucl\'{e}aire et Subnucl\'{e}aire,
Universit\'{e} de Mons,
UMONS Research Institute for Complex Systems,
Place du Parc 20, 7000 Mons, Belgium}

\author{\surname{Vincent} Mathieu \orcidlink{0000-0003-4955-3311}}
\email[E-mail: ]{vmathieu@ub.edu}
\affiliation{Departament de F\'isica Qu\`antica i Astrof\'isica and Institut de Ci\`encies del Cosmos, Universitat de Barcelona, E08028, Spain.}

\begin{abstract}
Glueballs are colour-singlet bound states built from gluons alone. They are an unavoidable consequence of the non-Abelian structure of Quantum Chromodynamics (QCD), and, in the pure Yang--Mills limit, they are the only physical excitations of the theory. The present article reviews what is known about them. After establishing which spin, parity and charge-conjugation quantum numbers two- and three-gluon states can carry, the pure-gauge spectrum is surveyed. Each of the main theoretical approaches and their respective conclusions are briefly presented. They include lattice QCD, constituent-gluon and Coulomb-gauge models, functional Dyson--Schwinger and Bethe--Salpeter equations, holographic models, and QCD sum rules. Particular attention is paid to the scale-setting ambiguity that limits how precisely a quenched glueball mass can be converted into physical units. The discussion then turns to full QCD. After discussing the meaning of a glueball assignment in this context, unquenching effects and questions related to glueball--$q\bar{q}$ mixing are addressed. The large-$N_c$ counting that underpins the mixing picture, as well as mass-matrix and effective-Lagrangian treatments of mixing are presented. The selection rules and dynamical mechanisms that shape glueball decays are also discussed. The gluon-rich production mechanisms used experimentally are finally reviewed. The candidates are assessed sector by sector: $f_0(1370)/f_0(1500)/f_0(1710)$ and the competing interpretations of the scalar sector, $\eta(1405)/\eta(1475)$ and $X(2370)$ in the pseudoscalar sector, the crowded tensor region, and the essentially unexplored $C=-1$ sector and its connection to the Odderon. Outlooks on the programmes that could help in validating some of these candidates or identifying others are reviewed as a conclusion.
\end{abstract}
\keywords{Exotic hadrons; Mesons; Glueballs; Non-Abelian gauge theory; Lattice QCD; Pomeron; Odderon; Hadron spectroscopy}

\maketitle

\tableofcontents

\newpage 

\section{Introduction}
\label{sec:intro}

Quantum Chromodynamics (QCD) is the gauge theory of the strong interaction, based on the non-Abelian colour group $SU(3)$. Its defining structural difference from Quantum Electrodynamics is that its gauge bosons themselves carry colour charge: gluons transform in the colour octet, whereas the photon is electrically neutral. Gluons therefore couple directly to one another through the three- and four-gluon vertices contained in the Yang-Mills Lagrangian
\begin{equation}
    \mathcal{L}_{\text{YM}} = -\tfrac{1}{4} F^{a}_{\mu\nu} F^{a\,\mu\nu},
    \qquad
    F^{a}_{\mu\nu} = \partial_\mu A^a_\nu - \partial_\nu A^a_\mu + g f^{abc} A^b_\mu A^c_\nu .
    \label{eq:ym}
\end{equation}
Because the gluon field is self-interacting, there is no reason for gluonic field configurations to require valence quarks in order to bind. Colour-singlet bound states made of gauge fields alone should therefore exist. These states are referred to as glueballs. This conjecture dates to the early 1970s, in the works of Fritzsch, Gell-Mann and Minkowski~\cite{frit72,frit73,frit75}. It predates by several years the experimental confirmation of the gluon itself in three-jet events at PETRA, whose history is reviewed in the companion article by Amsler \cite{amsl25}.

Glueballs are most easily studied in pure Yang--Mills theory, also referred to as quenched QCD. In this framework, all quark fields are removed from the QCD Lagrangian, leaving only the Yang-Mills contribution \eqref{eq:ym}. On the lattice, this theory is found to confine, meaning that its asymptotic states are colour-singlet bound-states of its coloured constituents. As this theory contains no quarks, its spectrum therefore consists entirely of glueballs\footnote{Note that whatever the lightest quenched glueball state is, its mass is the mass gap of $SU(3)$ Yang-Mills theory, generated from a classically scale-invariant action by dimensional transmutation, with $\Lambda_{\text{QCD}}$ as the only scale. Glueballs are thus not an exotic curiosity appended to QCD but the direct spectroscopic expression of its mass gap.}. Therefore, quenched QCD provides an ideal playground for assessing the main properties of glueballs from theory. This topic is discussed in \Cref{sec:pure_gauge}. This clear picture becomes significantly more complicated once quarks are reintroduced. Beyond rising technical complications, this change also affects the very meaning of a glueball assignment, a subject discussed in \Cref{sec:mix}. The remainder of \Cref{sec:full_qcd} theoretically introduces dynamical quarks, mixing and decay. Finally, \Cref{sec:production} describes the production mechanisms used experimentally, while \Cref{sec:experiment} assesses the candidates. \Cref{sec:outlook} concludes with an outlook on future developments, including future experimental programmes.

The main messages of this review are summarized in the following points. They are intended to provide a realistic state-of-the-art overview of the expected properties of glueballs. Each of these points is discussed in greater detail in the relevant sections, with further details given in the cited references.
\begin{itemize}[leftmargin=1.5em,rightmargin=3em,itemsep=-0.1em,topsep=0.3em]
    \item Two gluons can only form $C=+1$ states, and cannot have $J=1$. Glueballs with odd charge-conjugation require at least three gluons and, consequently, are much heavier.
    \item Glueballs have been studied theoretically (especially using lattice QCD calculations) for three decades, but have not been unambiguously identified in experiment.
    \item In quenched QCD, the scalar glueball is the ground state, with a mass of $3.40(2)$ in units of $\sqrt{\sigma}$. Converting this to physical units introduces an irreducible $15\%$ ambiguity.
    \item In full QCD, the gluonic content is an interpretive label. The true observables, which are measured and govern hadron phenomenology, are the pole spectrum and its couplings. Establishing a glueball component in any physical state requires experiment, amplitude analysis and theory to be confronted together. 
    \item Glueballs with non-exotic quantum numbers mix with $q\bar{q}$ isoscalars. Consequently, no physical resonance is a pure glueball.
    \item The leading candidates are currently $f_0(1710)$ and $X(2370)$ in the scalar and pseudoscalar sector respectively, but neither assignment is settled.
\end{itemize}

As glueballs, and hadron phenomenology more broadly, deserve considerably more space than can be devoted to it in this review, the present work is necessarily not exhaustive. To complement the discussion, we refer the reader to the following companion works.
\begin{itemize}[leftmargin=1.5em,rightmargin=3em,itemsep=-0.1em,topsep=0.3em]
    \item On glueballs themselves, earlier dedicated reviews of glueball physics include Refs.~\cite{math09,cred09,klem07,ochs13,llan21}.
    \item On the spectroscopy of light mesons, we refer to~\cite{pela25}. This article provides the $q\bar{q}$ baseline against which any supernumerary state must be counted, as well as the dispersive machinery needed to extract poles from data.
    \item Conventional constituent quark models are reviewed by Entem \emph{et al.} in \cite{ente25}. Their application to glueballs is discussed in~\Cref{sec:constituent}.
    \item Hadronic molecules and multiquark states are addressed in~\cite{hanh25}. As these are among the principal competing interpretations of candidates that do not fit the expected $q\bar{q}$ nonet, they deserve more consideration than can be given to them in the present work. 
    \item Along similar lines, hybrid mesons, in which an excited gluonic field is bound to a $q\bar{q}$ pair, are reviewed in ~\cite{dude26}. These states share the same non-perturbative gluon dynamics and are, in practice, among the closest neighbours of glueballs in the hadron spectrum.
    \item Regge theory is discussed in Sec.~\Cref{sec:regge}, as it supplies the framework in which high-spin glueballs organise themselves into trajectories. The leading $C=+1$ glueball trajectory is often identified with the Pomeron, while its $C=-1$ counterpart is related to the Odderon (see \Cref{sec:oddballs}). Regge theory is discussed in depth in~\cite{winn25}.
\end{itemize}

\section{The pure-gauge spectrum}
\label{sec:pure_gauge}

Removing the quark fields from QCD leaves the pure $SU(3)$ Yang-Mills theory of Eq.~\eqref{eq:ym}. This quenched limit is the natural first laboratory for the study of glueballs. These are stable there and do not mix with $q\bar{q}$ by construction. The only scale is the one the theory generates itself. In the following, before any dynamics, the quantum numbers accessible to quenched glueballs are addressed by resorting to group theory arguments. Then, several theoretical methods used to access properties of quenched glueballs are discussed. 

\subsection{Quantum numbers and selection rules}
\label{sec:quantum}

As required by confinement, a physical hadron must be a colour singlet. For two gluons, each transforming with the adjoint ($\mathbf{8}$) representation under colour transformation, the product decomposes as
\begin{equation}
    \mathbf{8}\otimes\mathbf{8} = \mathbf{1}\oplus\mathbf{8}_A\oplus\mathbf{8}_S\oplus\mathbf{10}\oplus\overline{\mathbf{10}}\oplus\mathbf{27}.
    \label{eq:colour2}
\end{equation}
The occurrence of a singlet component indicates that a physical colourless state can be built out of two gluons. The corresponding colour wave function,  $\delta^{ab}/\sqrt{8}$, is symmetric under interchange of the two gluons. For three gluons, two independent singlets are obtained,
\begin{equation}
    \mathbf{8}\otimes\mathbf{8}\otimes\mathbf{8} \supset \mathbf{1}\oplus\mathbf{1}.
    \label{eq:colour3}
\end{equation}
They are built on the antisymmetric structure constants $f^{abc}$ and the symmetric tensor $d^{abc}$, respectively. Due to this doubling, three-gluon states come in both charge-conjugation parities.

Apart from colour, gluons are massless vector particles and therefore genuinely possess only two physical polarisation states, labelled with a helicity quantum number $\lambda = \pm 1$. Any construction that assigns a gluon three spin projections, as a massive spin-1 particle, would have introduced degrees of freedom the gauge field does not possess. The spectra obtained with the most reliable techniques able to study quenched QCD turn out to favour the use of two helicity projections. A detailed comparison between two and three spin projections for the gluon is deferred to \Cref{sec:constituent}. 

For two-gluon glueballs, imposing Bose symmetry on the two-particle helicity basis of Jacob and Wick~\cite{jaco59} and taking into account the symmetric colour singlet constructed before, one finds that two-gluon states necessarily have
\begin{equation}
    C = +1,
\end{equation}
and 
\begin{equation}
    J^{PC} = 0^{++},\ 0^{-+},\ 2^{++},\ 2^{-+},\ 3^{++},\ 4^{++},\ 4^{-+},\,\ldots
    \label{eq:2gjpc}
\end{equation}
A few features of this list deserve emphasis. First, all two-gluon glueballs have positive charge conjugation. As already mentioned, the lowest glueball with $C=-1$ must necessarily contain at least three gluons. Second, $J=1$ is absent. This is a direct consequence of the Landau-Yang theorem, the familiar statement that a spin-$1$ particle cannot decay into two photons. If applied here to two massless vectors in a symmetric colour state, it tells that no $J=1$ state can be built. It is the reason the lightest spin-$1$ glueball (the $1^{+-}$) requires three gluons and lies far above the scalar. Third, no $J^{PC}$ in Eq.~\eqref{eq:2gjpc} is exotic. For a $q\bar{q}$ bound state, $P = (-1)^{L+1}$ and $C = (-1)^{L+S}$, which forbids the following $J^{PC}$ combinations,
\begin{equation}
    J^{PC} = 0^{--}, 0^{+-}, 1^{-+}, 2^{+-}, 3^{-+},\,\ldots
    \label{eq:exotic}
\end{equation}
The list from Eq.~\eqref{eq:2gjpc} never overlaps the above one: each $J^{PC}$ quantum numbers accessible to a two-gluon glueball is also accessible to a $q\bar{q}$ pair. This property strongly complicates the passage to full QCD, but this topic is deferred to \Cref{sec:mix}.

For three-gluon glueballs, combining both colour singlets from Eq.~\eqref{eq:colour3} with the appropriate helicity wave function to satisfy Bose symmetry, one obtains states of both charge conjugations. The $f^{abc}$ (antisymmetric) singlet yields $C = -1$ and the corresponding states are  conventionally called oddballs. The $d^{abc}$ (symmetric) singlet yields $C = +1$.

Oddballs are the interesting ones, because they include quantum numbers that are unavailable to a $q\bar{q}$ pair. Of these, the scalar $0^{\pm-}$ requires four gluonic operators in a symmetric configuration~\cite{Boulanger:2008aj} or three in a very non-symmetric one~\cite{chev25}. These are thus expected to be very heavy. However the quantum numbers $1^{\pm -}$ and $3^{\pm -}$ would be the lowest three-gluons states~\cite{Boulanger:2008aj}. Although they sit above $3$~GeV and decay to high-multiplicity final states. This renders their experimental study and the corresponding partial wave analysis significantly more intricate. The exotic combination $1^{-+}$ also belongs to the three-gluon $C=+1$ sector, but these quantum numbers are those of the lightest hybrid-meson candidates, whose mass is below 2 GeV~\cite{dude26}. For that reason, the observation of a light isoscalar $1^{-+}$ state (see \Cref{sec:pseudoscalar}) is commonly interpreted in terms of hybrids rather than glueballs. A similar issue also affects oddballs. The exotic quantum numbers they have access to are also accessible to tetraquark and molecular configurations. Consequently, the identification of a resonance with $J^{PC} =0^{+-}$ or $2^{+-}$ guarantees only that this state is not a simple $q\bar{q}$ meson and does not identify it as a glueball. \Cref{tab:jpc} summarises the state counting for two- and three-gluon glueballs. 

\begin{table}[t]
    \centering
    \caption{Charge conjugation and representative $J^{PC}$ available to glueballs built from two and three constituent gluons, and their status relative to the $q\bar{q}$ quark model. A dagger indicates quantum numbers forbidden to $q\bar{q}$. Only three-gluon configurations with $C=-1$ give access to these quantum numbers. For three-gluon glueballs, \cite{chev25} indicates that $0^{\pm-}$ glueballs can also be constructed but should be repelled to higher energies. \justifying}
    \label{tab:jpc}
    \vspace{2mm}
    \begin{tabular}{llll}
        \hline\hline
        Constituents\hspace{1cm} & Colour singlet\hspace{1cm} & $C$\hspace{1cm} & Representative $J^{PC}$ \\
        \hline
        $gg$   & $\delta^{ab}$ & $+1$ & $0^{++},\,0^{-+},\,2^{++},\,2^{-+},\,3^{++},\,4^{++},\,\ldots$ \\[2pt]
        $ggg$  & $d^{abc}$     & $+1$ & $0^{++}$,\,$0^{-+}$,\,$1^{++}$,\,$1^{-+}\,^{\dagger}$,\,$2^{++}$,\,$2^{-+}$\,\ldots \\[2pt]
        $ggg$  & $f^{abc}$     & $-1$ & $1^{+-}$,\,$1^{--}$,\,$2^{+-}\,^{\dagger}$,\,$2^{--}$,\,$3^{+-}$,\,$3^{--}$,\,\ldots \\
        \hline\hline
    \end{tabular}
\end{table}

\subsection{Lattice QCD}
\label{sec:lattice}

Lattice QCD discretises Euclidean spacetime on a hypercubic grid of spacing $a$ in a finite four-volume. This regulates the theory in both the ultraviolet (momenta above $\pi/a$ are forbidden) and the infrared (wavelengths beyond the box size $L$ are forbidden). Path integrals are then evaluated by Monte Carlo importance sampling, and physical results are recovered by extrapolating to infinite volume and to the continuum. It is the only method in this article that is systematically improvable: every approximation it makes can, in principle, be reduced by more computing.

Glueball masses are extracted from Euclidean correlators of gauge-invariant operators built from closed Wilson loops, projected onto the irreducible representations $A_1, A_2, E, T_1, T_2$ of the cubic rotation group and onto definite $P$ and $C$. Because these representations are not the continuum $SO(3)$ irreps, spin assignments must be reconstructed by identifying degenerate multiplets across representations (a step that becomes progressively less reliable as one climbs the spectrum). The practical difficulty of glueball calculations is that gluonic correlators are exceptionally noisy: the signal falls exponentially with the glueball mass while the statistical error falls only as the square root of the number of configurations. Anisotropic lattices, with a finer temporal than spatial spacing, together with variational bases of smeared operators, are what made the modern calculations feasible \cite{morn99,chen06,meye05,athe20}.

The results of the principal quenched calculations are collected in physical units in \Cref{tab:spectrum}. The qualitative picture has been stable for twenty-five years and is not in dispute.
\begin{itemize}[itemsep=-0em,]
    \item The scalar $0^{++}$ is the ground state and hence the Yang--Mills mass gap. In physical units, it is predicted to lie about $1.5$ to $1.8$~GeV. This uncertainty is discussed below.
    \item Then comes the tensor glueball $2^{++}$ which lies roughly $40\%$ above the lowest scalar glueball.
    \item The pseudoscalar $0^{-+}$ is close to, and slightly above, the tensor. It is the third lowest quenched glueballs.
    \item The entire $C=-1$ sector begins near twice the scalar mass.
\end{itemize}

\begin{figure}[t]
    \centering
    \includegraphics[width=0.92\linewidth]{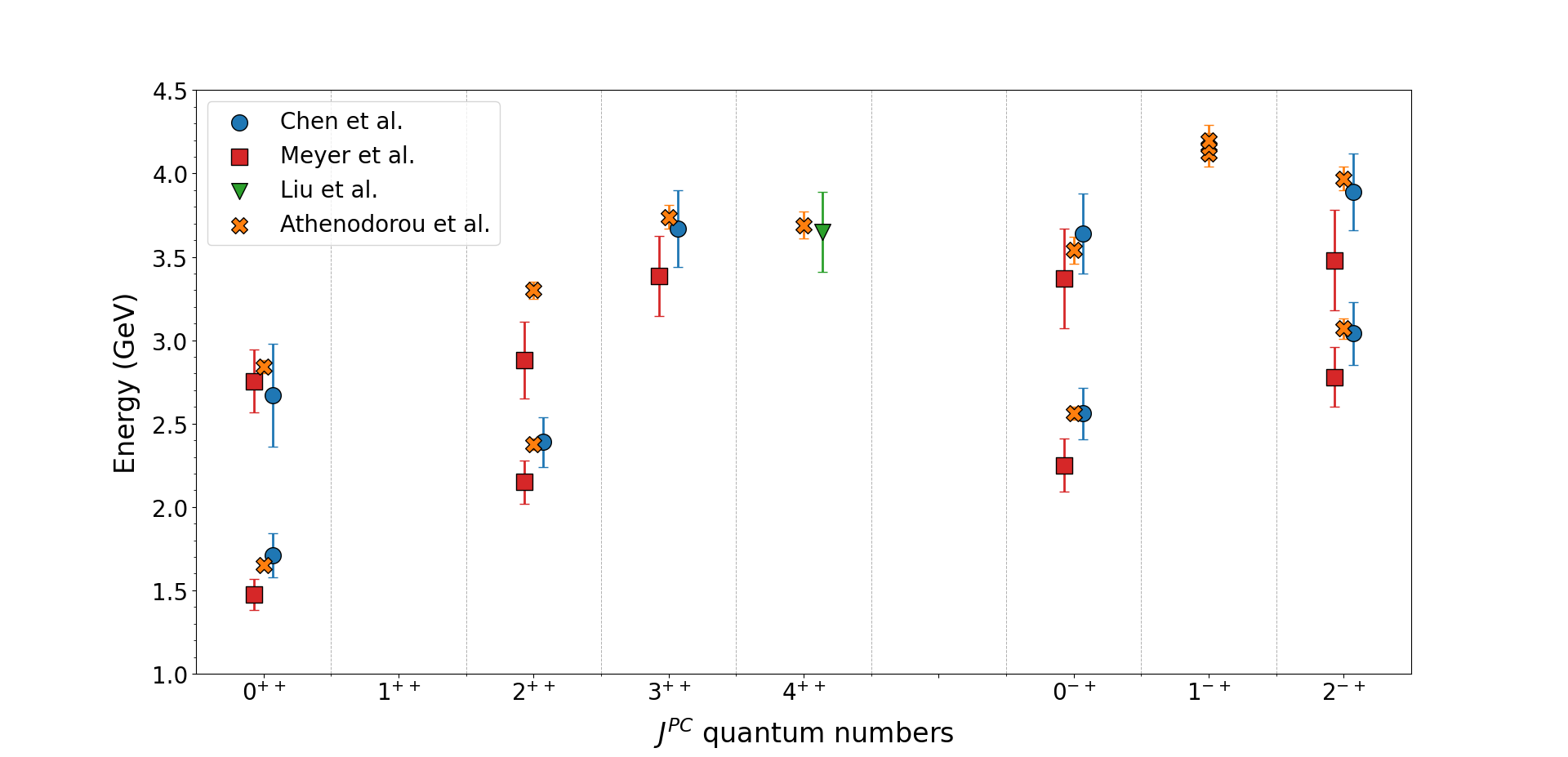}
    \includegraphics[width=0.92\linewidth]{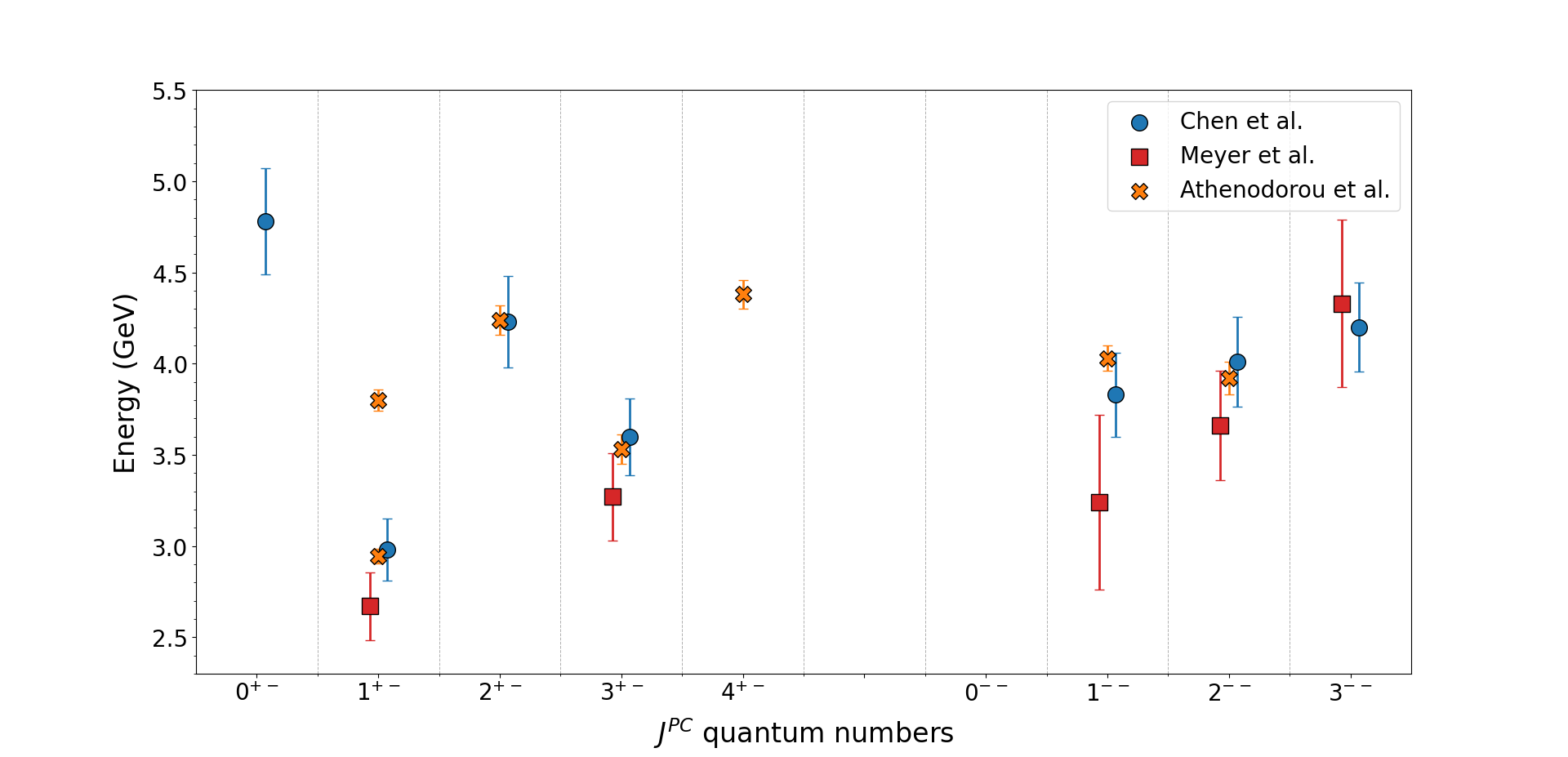}
    \caption{Quenched lattice glueball spectrum in the positive (top) and negative (bottom) charge-conjugation sectors, compiled from Refs.~\cite{morn99,chen06,meye05,athe20}. Note that the lightest $C=-1$ state is the $1^{+-}$, near $2.9$~GeV, and that the lightest $1^{--}$ lies almost a GeV higher. Vertical scales are subject to the scale-setting ambiguity discussed in \Cref{sec:scale}. \justifying}
    \label{fig:GB_spectrum_C}
\end{figure}

\subsubsection{The scale setting ambiguity}
\label{sec:scale}

It is common to see the quenched scalar glueball quoted as $1.7$~GeV with an uncertainty of a few tens of MeV. That precision is not real, and the reason is worth stating plainly because it propagates into every phenomenological analysis downstream. A pure Yang--Mills theory has one scale, and nothing in it is measured. To express a glueball mass in GeV one must borrow a scale from somewhere, and the choices available do not agree. To take an example, Athenodorou and Teper \cite{athe20} were able to determine the following dimensionless ratio to high precision,
\begin{equation}
    \frac{M_{0^{++}}}{\sqrt{\sigma}} = 3.405(21)
    \label{eq:ratio}
\end{equation}
where $\sigma$ is the confining string tension. To obtain a glueball mass in GeV, the scale $\sigma$ has to be fixed. Setting it through the Sommer parameter $r_0 = 0.472(5)$~fm gives $\sqrt{\sigma} = 485(6)$~MeV and hence $M_{0^{++}} = 1.65$~GeV. Setting it instead through the slope of the observed mesonic Regge trajectories gives the traditional $\sqrt{\sigma}\simeq 440$~MeV and yields $M_{0^{++}} \simeq 1.50$~GeV. Calculations that fix the scale from quenched charmonium splittings or from $f_\pi$ land closer to $1.7$--$1.8$~GeV. Morningstar and Peardon \cite{morn99} were explicit on this point: they quoted $m(0^{++}) = 1.730(50)(80)$~GeV with the second error arising entirely from the uncertainty in $r_0^{-1}$ and cautioned against direct comparison with experiment. The honest summary is therefore
\begin{equation}
    M_{0^{++}}\big|_{\text{quenched}} \;\approx\; 1.5\text{--}1.8\ \text{GeV},
    \label{eq:honest}
\end{equation}
with the ratio Eq.~\eqref{eq:ratio} known far better than either endpoint. This is not a defect of the lattice calculations, but a statement that the quenched theory is not QCD. Asking for its spectrum in GeV is asking a question with no unique answer. Arguments which turn on a $50$~MeV coincidence between a lattice number and a resonance mass should be treated with caution.

\subsubsection{Regge trajectories and the Pomeron}
\label{sec:regge}

High-spin glueballs organise themselves into approximately linear trajectories in the Chew--Frautschi plane,
\begin{equation}
    J = \alpha_0 + \alpha' M^2.
    \label{eq:regge}
\end{equation}
This is the spectroscopic signature of a rotating relativistic string, here a closed loop of chromoelectric flux rather than the open string appropriate to $q\bar{q}$ mesons. Using lattice QCD, Meyer and Teper \cite{meye05b} computed the lightest $J = 0, 2, 4, 6$ glueballs in the continuum limit and fitted the leading trajectory, obtaining
\begin{equation}
    \alpha(t) = 0.93(24) + 0.28(2)\,\alpha'_R\, t ,
    \label{eq:meyerteper}
\end{equation}
where $\alpha'_R \approx 0.9$~GeV$^{-2}$ is the slope of the ordinary mesonic trajectories. Overall, the glueball slope is roughly at $0.25$~GeV$^{-2}$. The result is expressed as a ratio because, the glueball trajectory being closed-string-like, its slope is expected to be a fraction of the open-string mesonic slope rather than to obey the open-string relation $\alpha'_{\text{open}} = 1/(2\pi\sigma) \approx 0.8$~GeV$^{-2}$.

This trajectory invites comparison with the soft Pomeron of high-energy diffractive scattering, whose Donnachie--Landshoff parameterisation has $\alpha_P(0) \simeq 1.08$ and $\alpha'_P \simeq 0.25$~GeV$^{-2}$ \cite{winn25}. In QCD, the Pomeron is understood as a color-singlet multi-gluon exchange that dominates high-energy diffractive processes, making its underlying Regge trajectory the natural home for glueball states. The slopes agree well. The intercepts are compatible, but, given that the lattice determination carries an uncertainty of $\pm0.24$, this comparison should be understood more as a consistency check than as a measurement. The lattice intercept of $0.93(24)$ neither confirms nor excludes $1.08$ and claims in the literature that the lattice reproduces the soft Pomeron intercept to a few percent should be read with this in mind. What the lattice does establish robustly is the qualitative statement that the leading $C=+1$ glueball trajectory has an intercept close to unity and a slope close to the phenomenological Pomeron slope, which is a non-trivial success of the identification of the Pomeron with glueball exchange. Nevertheless, two caveats are worth mentioning. First, the fit rests on very few states. No lattice calculation has reliably determined a glueball beyond $J=4$ (Athenodorou and Teper state explicitly that their data do not permit it). Second, trajectories built on the lowest few spins are not obviously in the asymptotic regime where the string picture applies.

\subsection{Constituent gluon models}
\label{sec:constituent}

Constituent models describe a glueball as a bound state of a small, fixed number of effective gluons interacting through a potential. They cannot compete with the lattice on precision, but they supply something the lattice does not: an interpretation. They say what a glueball is made of, why the levels appear in the order they do, and how many states to expect. A parallel pedagogical treatment for $q\bar{q}$ and $qqq$ systems is given by Entem \emph{et al.} in~\cite{ente25}. The next sections discuss the main constituent frameworks used to study glueballs, as well as the properties that should be attributed to the effective gluon, including its mass and its spin.

\subsubsection{The bag models}

The earliest treatment applied the MIT bag model to gluons \cite{jaff76,carl81,carl84}. Massless gluon fields are confined to a spherical cavity of radius $R$ held against an external vacuum pressure $B$. Low-lying glueballs are built by populating the transverse-electric (TE) and transverse-magnetic (TM) cavity modes. The lowest configuration $(\text{TE})^2$ produces a $0^{++}$ and a $2^{++}$, while $(\text{TE})(\text{TM})$ produces a $0^{-+}$. The model therefore reproduces the ordering seen on the lattice, which was a genuine early success, but its absolute scale comes out low, around $1$~GeV for the scalar. It is no longer used in quantitative spectroscopy: the sharp cavity wall breaks translational invariance and Lorentz covariance, the model has no dynamical chiral symmetry breaking, and the centre-of-mass motion must be removed by hand. It has been superseded by relativistic potential models and by the lattice.

\subsubsection{Dynamical mass generation and the constituent gluon}

Although the gluon is massless in all QCD Lagrangians (and remains massless to all orders in perturbation theory by gauge invariance), its propagator saturates in the infrared. This is understood as a non-perturbative realisation of the Schwinger mechanism, in which the vacuum polarisation develops a massless pole and the gauge boson acquires an effective mass without any breaking of gauge symmetry. The mechanism was first proposed by Cornwall \cite{corn82}, while its modern formulation in terms of massless bound-state excitations in the vertex is due to Aguilar, Ib\'a\~nez, Mathieu and Papavassiliou \cite{agui12}. The subsequent literature is reviewed in Refs.~\cite{agui16,papa22}. Lattice and functional determinations of the gluon propagator consistently indicate a dynamically generated mass
\begin{equation}
    m_g \sim 500\text{--}800\ \text{MeV}.
    \label{eq:mg}
\end{equation}
This property serves as an empirical justification for treating the gluon as a massive effective degree of freedom in constituent models, as detailed in the next sections. It also regulates the long-distance behaviour of one-gluon exchange.

\subsubsection{Semi-relativistic potential models and the gluon spin}

Modern constituent approaches replace the cavity from bag models with a two-body Hamiltonian of the semi-relativistic (spinless Salpeter) type,
\begin{equation}
    H = \sum_{i=1}^{n}\sqrt{\boldsymbol{p}_i^2 + m_g^2} \;+\; \sigma\, r \;+\; V_{\text{OGE}} + \ldots.
    \label{eq:hamiltonian}
\end{equation}
One recognizes the dynamical gluon mass of Eq.~\eqref{eq:mg}, a linear confining term, and the short-range potential deduced from one-gluon exchange \cite{corn83,brau04,math08b}. The delicate question is what spin structure to give the constituent gluon.

Treating the gluon as a massive spin-1 particle, with three polarisation states, is the naive choice and it fails in a diagnosable way: it generates states that pure-gauge lattice calculations do not contain. The clearest example is a low-lying two-gluon $1^{-+}$, which is excluded from the true spectrum by the Landau--Yang argument of \Cref{sec:quantum}. The extra states are an artefact of having given a massless gauge field a longitudinal polarisation it does not possess.

The alternative is to build states directly in the helicity basis \cite{jaco59}, restricting each gluon to $\lambda = \pm 1$ \cite{math08,boul08,chev25}. The unphysical projections then never enter. For two gluons the helicity formalism reproduces both the state counting and the level ordering of the quenched lattice, with no spurious $1^{-+}$. Chevalier and Mathieu \cite{chev25} extended the construction to three gluons, building totally symmetric three-particle helicity wave functions and computing the $C=-1$ spectrum; their two-gluon results agree well with the lattice, while their three-gluon spectrum broadly follows the lattice oddballs but predicts additional low-lying states not seen there, which remains to be understood. In this approach, the gluon is massless and its dynamical mass is interpreted as the expected value of its kinetic energy, $m_g^2 \sim \langle \sqrt{\bm p^2} \rangle$. 

\subsubsection{Coulomb gauge QCD} 

A closely related line of work quantises QCD in Coulomb gauge, where unphysical gauge and scalar degrees of freedom are eliminated at the outset and the surviving gluonic degrees of freedom are transverse by construction \cite{szcz96,szcz03,llan02,llan06}. This yields a many-body Hamiltonian that can be treated variationally. Coulomb-gauge and helicity-based constituent models share the essential feature that they work only within the physical transverse subspace, and both reproduce the lattice hierarchy without spurious states. They are not, however, equivalent frameworks, and they differ in their treatment of the confining kernel and of gluon self-energy effects.

\subsection{Functional methods}
\label{sec:functional}

Dyson-Schwinger equations (DSE) for QCD's correlation functions, combined with Bethe-Salpeter equations (BSE) for bound states, provide a continuum field-theoretic route to the spectrum with no lattice discretisation. A glueball appears as a pole in a four-point function, obtained by solving the homogeneous BSE for the bound-state amplitude $\Gamma(p;P)$,
\begin{equation}
    \Gamma(p; P) = \int\!\frac{\diff^4 k}{(2\pi)^4}\, K(p,k;P)\, D(k + P/2)\, D(k - P/2)\, \Gamma(k;P),
    \label{eq:bse}
\end{equation}
where $D$ is the fully dressed gluon propagator and $K$ the interaction kernel, both determined from the DSE tower.

Huber, Fischer and Sanchis-Alepuz \cite{hube20,hube21,hube25} have carried this programme to a self-contained truncation in which the propagators and vertices are computed within the same framework, with no external modelling and no fitted parameters beyond the overall scale. They obtain $M_{0^{++}} = 1.850(130)$~GeV and $M_{0^{-+}} = 2.580(180)$~GeV, and have extended the calculation to $J\le 4$, finding $M_{2^{++}} = 2.610(180)$~GeV. A comparison with the lattice QCD spectrum is displayed in \Cref{tab:spectrum}. For the scalar and pseudoscalar, the masses from DSE/BSE show agreement with the lattice at the $5\text{--}10\%$ level. However, relative to the lattice, the tensor comes out somewhat high, and the $2^{-+}$ somewhat low.

The characteristic limitation of the method is truncation. The DSE for an $n$-point function involves $(n+1)$- and $(n+2)$-point functions, so the tower must be closed by hand, and the reliability of the result depends on where and how this is done. The recent inclusion of two-loop diagrams in the Bethe-Salpeter kernels produces only marginal shifts in the three lightest glueballs \cite{hube25}, which is evidence, though not proof, of convergence. 

\begin{table}[t]
    \centering
    \caption{Ground-state glueball masses in GeV from three quenched lattice calculations and from functional (DSE/BSE) methods. To make the comparison meaningful, all entries are taken from the compilation of Ref.~\cite{hube21}, in which the lattice results of Refs.~\cite{morn99,chen06,athe20} have been placed on a common scale using $r_0^{-1} = 418(5)$~MeV. The quoted errors are those of the compilation and do not include the scale-setting uncertainty discussed in \Cref{sec:scale}. This uncertainty is common to all columns and is of order $10\text{--}15\%$. \justifying}
    \label{tab:spectrum}
    \vspace{2mm}
    \begin{tabular}{lrrrr}
        \hline\hline
        $J^{PC}$\hspace{1cm} & \hspace{1cm}Ref.~\cite{morn99} & \hspace{1cm}Ref.~\cite{chen06} & \hspace{1cm}Ref.~\cite{athe20} & \hspace{1cm}DSE/BSE \cite{hube20,hube21} \\
        \hline
        $0^{++}$ & $1.760(50)$ & $1.740(60)$ & $1.651(23)$ & $1.850(130)$ \\
        $2^{++}$ & $2.447(25)$ & $2.440(50)$ & $2.376(32)$ & $2.610(180)$ \\
        $0^{-+}$ & $2.640(40)$ & $2.610(50)$ & $2.600(40)$ & $2.580(180)$ \\
        $2^{-+}$ & $3.160(31)$ & $3.100(60)$ & $3.070(60)$ & $2.740(140)$ \\
        $3^{++}$ & $3.760(40)$ & $3.740(60)$ & $3.740(70)$ & $3.370(50)$ \\
        $4^{++}$ & --- & --- & $3.690(80)$ & $4.140(30)$ \\
        \hline
        $1^{+-}$ & --- & --- & $2.942(41)$ & --- \\
        $1^{--}$ & --- & --- & $\sim 3.840(70)$ & --- \\
        \hline\hline
    \end{tabular}
\end{table}

\subsection{Holographic models}
\label{sec:holography}

Gauge/gravity duality provides a further continuum approach. In its original form, glueballs of a strongly coupled large-$N_c$ gauge theory are dual to supergravity modes in a five-dimensional warped geometry, and their masses follow from solving the linearised bulk equations of motion subject to appropriate boundary conditions \cite{csak99,brow00}. Bottom-up ``AdS/QCD'' constructions (including hard-wall, soft-wall and graviton soft-wall models) modify the geometry or add a dilaton profile so as to build in confinement and reproduce linear Regge trajectories \cite{bosc06,cola07,fork08,vent17,rina20,rina21}.

These models are not first-principles QCD: the dual of QCD is not known, and the models involve choices of geometry and of the operator-field dictionary. Their value is that they generate an entire spectrum, including the correct Regge systematics, from very few parameters, and that they make predictions for quantities such as decay rates that are difficult to obtain elsewhere. The top-down Witten-Sakai-Sugimoto model is the most predictive of these and is used in \Cref{sec:scalar} to confront glueball decay patterns with data \cite{brun15}.

\subsection{QCD sum rules}
\label{sec:sumrules}

QCD sum rules relate hadronic properties directly to the QCD Lagrangian through the analytic structure of correlation functions \cite{shif79,cola01}. One studies the two-point function of a gluonic interpolating current,
\begin{equation}
    \Pi(q^2) = i\!\int\!\diff^4 x\, e^{i q\cdot x}\, \bra{0}\, T\{ J_G(x)\, J_G(0) \}\, \ket{0},
    \label{eq:correlator}
\end{equation}
with $J_S = \alpha_s F^a_{\mu\nu}F_a^{\mu\nu}$ for the scalar channel and $J_P = \alpha_s F^a_{\mu\nu}\tilde{F}_a^{\mu\nu}$ for the pseudoscalar, where $\tilde{F}^a_{\mu\nu} = \tfrac12 \epsilon_{\mu\nu\alpha\beta}F^{a\,\alpha\beta}$. At short distances $\Pi$ is computed via the operator product expansion, which separates perturbative contributions from vacuum condensates such as $\braket{\tfrac{\alpha_s}{\pi}F^2}$. On the hadronic side it is written as a dispersion integral over a resonance pole plus a continuum above a threshold $s_0$. Borel transformation suppresses the continuum and the higher condensates, and matching the two representations yields the mass and coupling.

The glueball channels are unusual in that the standard OPE is insufficient. Instantons, which are topological fluctuations of the gauge field, couple very strongly to $F^2$ and $F\tilde{F}$. For that reason, their direct contribution must be added explicitly. Forkel~\cite{fork05} showed that these direct-instanton contributions are attractive in the $0^{++}$ channel, pulling the extracted scalar mass down towards $1.2\text{--}1.5$~GeV, and repulsive in the $0^{-+}$ channel, pushing it up. They are therefore the mechanism responsible for the large scalar--pseudoscalar splitting, and their inclusion is what allows sum rules to satisfy the low-energy theorems in both channels simultaneously. These theorems, derived from QCD scale and axial anomalies, dictate the exact zero-momentum behaviour of $F^2$ and $F\tilde{F}$ correlators, serving as non-perturbative boundary conditions that sum rules must obey. Sum-rule determinations remain spread over a wider range than lattice ones, roughly $1.2\text{--}1.8$~GeV for the scalar depending on the treatment of instantons and of the continuum \cite{shif79,nari98,fork05}. The method is best regarded as providing a consistency constraint rather than a precision determination.

\subsection{Conclusion: comparing the methods}
\label{sec:synthesis}

The apparent agreement between each framework is less trivial than it looks, as all the frameworks described above encode non-perturbative physics in a different way.
\begin{itemize}
    \item On the lattice, it is not parametrised at all and is simply the result of the simulation.
    \item In constituent models the non-perturbative input is put in by hand as a spatial potential: a linear term $\sigma r$ for confinement and a dynamical gluon mass $m_g$ that both regulate one-gluon exchange and sets the constituent scale.
    \item In functional methods, it is encoded in momentum space, in the infrared behaviour of the dressed propagators and vertices. The three-dimensional Fourier transform of an infrared-enhanced kernel is a confining spatial potential, so the two descriptions are related by the same reduction that takes a four-dimensional Bethe-Salpeter equation to a three-dimensional Salpeter equation under an instantaneous approximation.
    \item In holographic models, non-perturbative physics is geometric, residing in the warp factor and the dilaton profile.
    \item In sum rules, non-perturbative physics enters analytically, through local condensates in the OPE and through non-local topological configurations.
\end{itemize}
Given how different these frameworks are, the convergence on a common picture proves meaningful:
\begin{equation}
    M_{0^{++}} < M_{2^{++}} \lesssim M_{0^{-+}} \ll M_{\text{oddballs}}.
\end{equation}
The methods agree that the scalar provides the mass gap, that the scalar-to-tensor ratio is about $1.4$, and that the lightest $C=-1$ state is near twice the scalar mass. On the other hand, where the methods show disagreement, these differences can be traced to identifiable ingredients, rather than being unexplained. For instance, the scalar glueball from sum-rule lies lower, probably due to the treatment of instantons, while the tensor from functional methods lies higher, probably due to kernel truncation.

Because it is the pivot on which the identification of experimental candidates turns, one property of the pure-gauge theory deserves a closing remark. In pure Yang-Mills, because there are no quarks for a glueball to decay into, every glueball is absolutely stable: $\Gamma_G \equiv 0$. Every question about glueball widths, branching ratios, decay systematics and experimental signatures therefore arises only when dynamical quarks are switched on. That is the subject of the next section.

\section{Glueballs in full QCD: mixing and decay}
\label{sec:full_qcd}

Switching on light dynamical quarks changes the problem qualitatively. Glueballs stop being eigenstates: they acquire widths, and they mix with the $q\bar{q}$ isoscalars that share their quantum numbers. Neither effect is a small perturbation, and both must be understood before any experimental candidate can be assessed.

\subsection{Mixing with isoscalar mesons}
\label{sec:mix}

If glueballs are easily studied in the quenched approximation, they have proved remarkably elusive in practice. This is because, in the real world, quarks are light and dynamical. Because the lowest-lying glueballs carry $J^{PC} = 0^{++}$, $2^{++}$ and $0^{-+}$, all of which are quantum numbers that ordinary isoscalar $q\bar{q}$ mesons also carry, nothing forbids mixing. As a result, any physical resonance in these channels is a superposition of the possible bare resonances,
\begin{equation}
    \ket{R} = a\ket{G} + b\ket{n\bar{n}} + c\ket{s\bar{s}} + \ldots
    \qquad \text{where } \ket{n\bar{n}} \equiv \tfrac{1}{\sqrt{2}}\big(\ket{u\bar{u}}+\ket{d\bar{d}}\big)
    \label{eq:superposition}
\end{equation}
The direct search for a glueball state $\ket{G}$ is thus replaced by the task of attributing values for the coefficients $a$, $b$  and $c$ to each physical resonance. That question cannot be answered by a single measurement. It requires a coupled-channel amplitude analysis of several production reactions and several decay modes. The experimental results are then compared against theoretical predictions for masses, mixing amplitudes, production strengths and branching ratios. Both steps carry model dependence.

Glueballs with quantum numbers forbidden to $q\bar{q}$ escape this problem, but they are not necessarily easier to tackle. As discussed in \Cref{sec:quantum}, they require at least three gluons and therefore lie well above $3$~GeV. In addition, if mixing with conventional mesons is forbidden, they are not protected against mixing with four-quark configurations or hadronic molecules. For these reasons, experimental efforts have so far focused primarily on the low-lying glueballs.

\subsection{The meaning of a glueball assignment}
\label{sec:whatmatters}

The possibility of mixing between glueballs and isoscalar mesons significantly affects the interpretation of a glueball assignment. The true observable, experimentally accessible, is the spectrum of hadron resonances. A correct analysis of a scattering experiment enables one to access positions and residues of poles in the complex energy plane, as well as their couplings to the channels through which they are produced and into which they decay. The gluonic content of a resonance is, by contrast, not an observable. There is no possible measurement that enables a direct access to the internal structure of any physical state. If a partial-wave analysis determines the existence of a pole at $(1733 - 75\,i)$~MeV, with given couplings to $\pi\pi$, $K\bar{K}$ and $\eta\eta$, it has determined everything that reaction theory can access and has not determined whether the state is a glueball. 

Actually, the weight of one component in a decomposition into basis states is not unique: not only it depends on the operator basis chosen, but the mixing between gluonic and quark bilinear operators is also scheme- and scale-dependent under renormalisation. Therefore, two analyses may legitimately assign different glueball fractions to the same physical pole without either being wrong. Nor does such a label carry much direct phenomenological consequence. Knowing that a given $f_0$ is predominantly gluonic changes no cross section and enters no nuclear force calculation, since these states are too heavy and too weakly coupled to nucleons to matter there. Even in high-energy diffraction processes, in which gluonic degrees of freedom demonstrably control hadronic observables through the Pomeron and the Odderon (\Cref{sec:regge,sec:oddballs}), the connection is to the Regge trajectory as a whole, rather than to the identification of any individual resonance.

Assigning labels to physical resonances is better understood as an organising principle rather than a measurement. It compresses a large number of otherwise unrelated observations into a few statements.
\begin{itemize}
    \item It explains why the scalar isoscalar sector is overpopulated relative to the quark model.
    \item It predicts relations among branching ratios (see the flavour blindness hypothesis for glueball decay and its modification by chiral suppression).
    \item It predicts a hierarchy of production strengths in gluon-rich versus gluon-poor reactions.
    \item It tells experimenters where in the spectrum to look for states not yet found.
    \item It ties the observed hadron spectrum back to the mass gap and the confinement mechanism of the underlying gauge theory
\end{itemize}
A glueball assignment is valuable not because it is a new piece of data but because it makes the existing data intelligible, what is by itself a real scientific goal. But, as explained above, this cannot be settled by any single measurement or single calculation. Establishing a glueball component requires
\begin{itemize}
    \item to extract a pole and its couplings from several reactions by amplitude analyses,
    \item to compute theoretical predictions for masses, mixing amplitudes, production rates and decay patterns (at least resorting to controlled approximations),
    \item to conduct a genuine confrontation of the two (so that the quantities being compared are defined the same way on both sides).
\end{itemize}
That is a collaborative undertaking spanning lattice practitioners, continuum theorists, amplitude analysts and several experimental collaborations. In the end, the current disagreements are a symptom of that collaboration being incomplete rather than of the data being insufficient, especially in the scalar sector, as discussed in \Cref{sec:scalar}.

\subsection{Insights from the large-\texorpdfstring{$N_c$}{Nc} expansion}
\label{sec:largeN}

Closing the digression on the meaning of a glueball assignment to a physical state, let us return to the description of glueballs in full QCD. The cleanest organising principle here is deduced from the large-$N_c$ expansion. Counting powers of $N_c$ in the standard way \cite{thoo74,witt79}, one finds the following scaling for the couplings relevant to glueball phenomenology,
\begin{equation}
\begin{aligned}
    \text{meson} \to \text{meson} + \text{meson}\quad &\sim\ 1/\sqrt{N_c},\\
    \text{glueball} \to \text{meson} + \text{meson}\quad &\sim\ 1/N_c,\\
    \bra{G} H \ket{q\bar{q}} \ \ (\text{mixing})\quad &\sim\ 1/\sqrt{N_c}.
\end{aligned}\label{eq:largeN}
\end{equation}
Glueball masses themselves are $O(1)$. Hadronic widths scale as the square of the couplings. Thus the glueball widths scale as $1/N_c^2$ against $1/N_c$ for ordinary mesons and glueballs are expected to be narrower than $q\bar{q}$ states of comparable mass. This expectation is worth holding onto when assessing candidates: a very broad state is not a natural glueball. This is one of the difficulties facing the identification of the scalar glueball with any single broad $f_0$ resonance. Second, and less comfortably, the mixing probability scales as $1/N_c$. Large-$N_c$ therefore predicts that glueball-$q\bar{q}$ mixing should be suppressed but not small. Everything in \Cref{sec:experiment} is a consequence of that.

\subsection{Unquenched lattice QCD}
\label{sec:unquenched}

Lattice simulations with dynamical sea quarks are far more demanding than quenched ones for glueballs specifically. The relevant correlators now include quark-disconnected diagrams, namely closed quark loops with no valence line connecting them to the source, which are numerically expensive and intrinsically noisy. Progress has consequently been slower than in most other areas of lattice spectroscopy \cite{mcne01,rich10,greg12,sun18,jian23}.

Three classes of results are established, concerning, respectively, the stability of the spectrum, the mixing amplitudes, and the production rates. The third class, concerning production rates, is deferred to \Cref{sec:radiative}, while the other two can already be discussed. Concerning the spectrum, including dynamical light quarks shifts the mass of the bare scalar glueball only modestly \cite{rich10,greg12,sun18}, so the quenched value remains a useful benchmark for the unmixed state. Gregory \emph{et al.} \cite{greg12} report masses for ten glueball states on $N_f=2+1$ configurations at $m_\pi = 360$~MeV, with results broadly consistent with the quenched spectrum, though they do not extract mixing fractions for physical resonances. Regarding mixing amplitudes, McNeile and Michael \cite{mcne01} pioneered the direct extraction of scalar glueball--$q\bar{q}$ mixing on the lattice. More recently, Jiang \emph{et al.} \cite{jian23} performed the first determination of \emph{pseudoscalar} $\eta$--glueball mixing in $N_f=2$ QCD at $m_\pi\approx 350$~MeV. They found a mixing angle $\vert{}\theta\vert{} = 3.46(46)^\circ$ and a mixing matrix element $\vert{}x\vert{} = 107(15)(2)$~MeV. A mixing angle of this size corresponds to a gluonic probability below $1\%$ in the $\eta$, consistent with the expectation that $U(1)_A$ anomaly dynamics generates the $\eta'$ mass rather than large mixing with a heavy $0^{-+}$ glueball.

Nevertheless, all these calculations still treat glueballs as stable states extracted from a variational basis, whereas the physical states are broad resonances lying above several thresholds. A rigorous treatment requires finite-volume L\"uscher-type analyses in coupled channels, which for scalar isoscalars has not yet been done, the $\pi\pi$, $K\bar{K}$, $\eta\eta$ and $4\pi$ channels being all open. This is the single largest gap between lattice glueball calculations and experiment.

\subsection{Mass-matrix mixing schemes}
\label{sec:mixing}

The most widely used phenomenological framework treats mixing through a mass matrix in a basis of unmixed states. In the scalar isoscalar sector, one takes the states from Eq.~\eqref{eq:superposition}, $\{\ket{n\bar{n}}, \ket{s\bar{s}}, \ket{G}\}$, and writes their squared mass matrix
\begin{equation}
    M^2 = \begin{pmatrix}
        M_{n\bar{n}}^2 & 0 & f \\
        0 & M_{s\bar{s}}^2 & \sqrt{2}\,f \\
        f & \sqrt{2}\,f & M_G^2
    \end{pmatrix},
    \label{eq:massmatrix}
\end{equation}
where $f = \bra{n\bar{n}}H_{\text{mix}}\ket{G}$ is the glueball-quarkonium mixing amplitude. The relative factor $\sqrt{2}$ between the $\ket{n\bar{n}}$ and $\ket{s\bar{s}}$ entries follows from the flavour blindness hypothesis of the gluonic coupling together with the normalisation of $\ket{n\bar{n}}$. Allowing for flavour-symmetry breaking, this factor is often replaced by a fitted parameter 
\begin{equation}
    r_s = \frac{\bra{s\bar{s}}H_{\text{mix}}\ket{G}}{\bra{n\bar{n}}H_{\text{mix}}\ket{G} \times \sqrt{2}}
\end{equation}
with $r_s < 1$. The three physical states are obtained by diagonalising the above $M^2$ matrix,
\begin{equation}
    \begin{pmatrix} \ket{R_1} \\ \ket{R_2} \\ \ket{R_3}\end{pmatrix}
    = \mathbf{U}\begin{pmatrix}\ket{n\bar{n}} \\ \ket{s\bar{s}} \\ \ket{G}\end{pmatrix}.
\end{equation}
The glueball content of the physical state $\ket{R_i}$ is given by $|U_{i3}|^2$.

Results obtained from such a construction should be read as indicative of gross composition rather than as percentage determinations. Above, the $M^2$ matrix is a Hermitian, energy-independent mass matrix, which therefore describes stable states. In reality, the physical resonances have widths of $100\text{--}400$~MeV, comparable to their splitting. A proper treatment requires a complex, energy-dependent matrix or a full coupled-channel amplitude. Finally, the input $M_G$, normally taken from quenched lattice, is, per \Cref{sec:scale}, uncertain at the $\pm 150$~MeV level. This uncertainty is larger than the splitting the fit is trying to resolve. 

\subsection{Effective Lagrangians and symmetry constraints}
\label{sec:eft}

A complementary approach to the above one constrains glueball couplings by symmetry alone. In extended linear sigma models, a scalar glueball field $\chi$ identified with the dilaton associated with the scale anomaly is coupled to the scalar and pseudoscalar meson nonets \cite{giac05,jano14}. Schematically,
\begin{equation}
    \mathcal{L} = \tfrac{1}{2}(\partial_\mu\chi)^2 - V_{\text{dil}}(\chi) + \mathcal{L}_{\text{mesons}}(\Phi) + g_{\chi\Phi}\,\chi\,\text{Tr}\!\left(\Phi^\dagger \Phi\right) + \ldots,
    \label{eq:eLSM}
\end{equation}
where $\Phi$ is the meson matrix and $V_{\text{dil}}$ is the logarithmic dilaton potential whose divergence reproduces the QCD trace anomaly. The virtue of this construction is that chiral symmetry and $SU(3)_f$ fix the ratios of many couplings. The small number of remaining parameters are fitted to well-measured meson decays, thereby yielding absolute predictions for glueball partial widths into $\pi\pi$, $K\bar{K}$, $\eta\eta$, $\eta\eta'$ and vector pairs. Janowski \emph{et al.} \cite{jano14} concluded on this basis that $f_0(1710)$ carries the dominant glueball component. Giacosa \emph{et al.} \cite{giac05}, with different assumptions about the scalar nonet, had earlier favoured $f_0(1500)$. At the end, the framework is a tool for turning an assignment into testable predictions rather than an independent determination of the assignment.

\subsection{Selection rules for glueball decays}
\label{sec:selection}

Let us move on from mixing to the analysis of glueball decay properties. Which final states a glueball can reach can be fixed before any dynamics. For a pair of neutral pseudoscalars in relative orbital angular momentum $L$, Bose symmetry and charge conjugation give $C = (-1)^L$ and parity gives $P = (-1)^L$. Therefore, a system of two pseudoscalars with $C=+1$ has even $L$ and positive parity. This impacts the decay channel accessible to the lowest bare glueballs.
\begin{itemize}[itemsep=-0em,topsep=0.6em]
    \item Scalar and tensor glueballs decay to two pseudoscalars in $L=0$ and $L=2$, respectively. The accessible channels are $\pi^+\pi^-$, $\pi^0\pi^0$, $K^+K^-$, $K^0_S K^0_S$, $\eta\eta$ and $\eta\eta'$. At higher masses, four-pion and vector-pair modes ($\rho\rho$, $\omega\omega$, $\phi\phi$) become also possible.
    \item As the decay of any glueball with positive charge conjugation into a pair of pseudoscalars requires $P=+1$, pseudoscalar glueballs cannot decay to two pseudoscalars at all. This is why the searches for $0^{-+}$ meson resonances are conducted in vector-pair modes and three-body channels, such as $K\bar{K}\pi$ (typically through $K^*\bar{K}$), $\eta\pi\pi$ or $\eta'\pi\pi$.
\end{itemize}
The absence of the two-pseudoscalar mode in the pseudoscalar sector is a real experimental handicap. It removes the cleanest, highest-statistics final states and forces reliance on multi-body partial-wave analyses.

\subsection{Flavour blindness and chiral suppression}
\label{sec:flavour}

Beyond selection rules, the pattern of branching ratios is the main dynamical handle on gluonic content. The naive expectation is flavour blindness: a pure glueball is a flavour singlet and couples equally to $u\bar{u}$, $d\bar{d}$ and $s\bar{s}$. This hypothesis gives, for an $SU(3)$-symmetric coupling and with the $\eta$ treated in the singlet-octet basis, the following ratios for branching fractions
\begin{equation}
    \pi\pi : K\bar{K} : \eta\eta \;=\; 3 : 4 : 1,
    \label{eq:flavourblind}
\end{equation}
with $\eta\eta'$ vanishing in the symmetric limit \cite{amsl95}. 

Any deviation from Eq.~\eqref{eq:flavourblind} requires an explanation. Two mechanisms can generate them. The first, and less interesting, is kinematic. Because of the hierarchy $m_\pi < m_K < m_\eta < m_{\eta'}$, phase space suppresses the heavier channels near threshold. For a state at $1.5$ to $1.7$~GeV the effect on $\eta\eta'$ in particular is large. This correction can be computed and removed. The second mechanism is dynamical. Chanowitz \cite{chan05} pointed out that a $0^{++}$ glueball couples to a $q\bar{q}$ pair through a vertex requiring a helicity flip. This causes the amplitude to be proportional to the quark mass and makes the decay to light quarks suppressed in the chiral limit. Concretely, this chiral suppression mechanism predicts an enhancement of $K\bar{K}$ and $\eta\eta$ over $\pi\pi$, in direct opposition to the prediction of Eq.~\eqref{eq:flavourblind}. It is the main theoretical argument in favour of identifying the scalar glueball with the kaon-rich $f_0(1710)$ rather than with the pion-rich $f_0(1500)$.

The strength of the chiral suppression effect has been disputed. Chanowitz's argument is a chirality argument at the quark level: the amplitude for $G\to q\bar{q}$ is proportional to the quark mass to all orders in perturbation theory, and with current quark masses ($m_s/m_{u,d}\sim 20$) the suppression is dramatic. Chao, He and Ma \cite{chao07} objected that, in a perturbative QCD treatment, the scalar glueball couples to two $q\bar{q}$ pairs rather than one, so that the suppression at the hadron level is governed by decay constants rather than by masses. This gives a much milder effect: $\mathcal{B}(\pi\pi)/\mathcal{B}(K\bar{K}) \sim f_\pi^4/f_K^4 \approx 0.48$. Chanowitz replied \cite{chan07} reaffirming the quark-level result. The disagreement has not been resolved, and since chiral suppression is load-bearing for the $f_0(1710)$ interpretation (\Cref{sec:scalar}), it should be treated as a genuine systematic uncertainty rather than a settled input.

\subsection{Radiative \texorpdfstring{$J/\psi$}{J/psi} decay rates from the lattice}
\label{sec:radiative}

A prediction that is both first-principles and directly comparable to experiment is the rate at which a glueball is produced in radiative $J/\psi$ decay. The lattice computes the relevant transition matrix element in the quenched or partially quenched theory. The results are
\begin{align}
    \mathcal{B}\big(J/\psi\to\gamma\, G_{0^{++}}\big) &= 3.8(9)\times 10^{-3} & \cite{gui13},\\
    \mathcal{B}\big(J/\psi\to\gamma\, G_{2^{++}}\big) &\approx 1.1\times 10^{-2} & \cite{yang13}, \label{eq:tensor_Jpsi}\\
    \mathcal{B}\big(J/\psi\to\gamma\, G_{0^{-+}}\big) &= 2.31(90)\times 10^{-4} & \cite{gui19},
\end{align}
building on the original calculation of Sexton, Vaccarino and Weingarten \cite{sext96}. These are large rates: the tensor prediction in particular is a substantial fraction of the total radiative width. It is one of the strongest reasons to expect that if tensor glueballs exist, radiative $J/\psi$ decay is where they will be found. They are also the sharpest quantitative targets available, since unlike masses they are not degraded by the scale-setting ambiguity in the same way. Nevertheless, as for masses, they are computed for stable glueballs in a quenched or partially quenched world. The physical production rate is distributed over the mixed physical resonances in proportion to their glueball content, which is exactly the quantity one is trying to measure. But used carefully, this circularity is an advantage: comparing measured radiative production strengths across the $f_0$ states, against a total fixed by the lattice, constrains the distribution of glueball content among them.

\section{Production mechanisms}
\label{sec:production}

As already approached at the end of the previous section, because glueballs are not distinguished by their quantum numbers, the experimental strategy rests instead on where they are produced. The logic is to enrich the sample in gluons and deplete it in valence quarks. The resulting data are then compared with what is obtained in gluon-poor reactions. Four mechanisms carry most of the weight, and a fifth is used as a veto.

\paragraph{Radiative quarkonium decay.} 

In $J/\psi$ radiative decay, $J/\psi\to\gamma gg$, the $c\bar{c}$ pair annihilates after emitting a hard photon, leaving a two-gluon system with $C=+1$ in a colour singlet. This is precisely the configuration from which a two-gluon glueball is built. The $J/\psi$ lies below the open-charm threshold and its decay to light quarks is OZI-suppressed, so the radiative branching fraction is large and the gluonic content of the recoiling system is high. The lattice rates of \Cref{sec:radiative} apply directly to this process. The decisive datasets come from BESII and BESIII at BEPC, which has accumulated more than $10^{10}$ $J/\psi$ events, with earlier contributions from MARK~III at SPEAR and CLEO/CLEO-c at CESR. Radiative $\psi(2S)$ and $\Upsilon$ decays offer the same mechanism at higher mass but with far smaller samples.

\paragraph{Central exclusive production.}

In double-Pomeron exchange, $pp \to p_{\text{fast}}\, X\, p_{\text{slow}}$, both protons survive and the centrally produced system $X$ is formed from the collision of two colourless, gluon-dominated exchanges. It therefore carries vacuum quantum numbers and is expected to be gluon-rich. The classic measurements are those of WA76, WA91 and WA102 at the CERN SPS, together with GAMS. The WA102 experiment established the empirical ``$\diff P_T$ filter'', in which selecting events with a small vector difference of the transverse momenta transferred at the two proton vertices suppresses ordinary $q\bar{q}$ production relative to glueball candidates. The mechanism is now accessible at much higher energies at the LHC, where LHCb, CMS and the TOTEM/ATLAS forward detectors can tag both protons. This programme is in its early stages for light-meson spectroscopy but is the natural successor to WA102.

\paragraph{Antiproton annihilation.}

Antiproton annihilation at rest or in flight on hydrogen, proceeding through protonium, produces multi-meson final states in a gluon-dense environment with strong constraints from the well-defined initial atomic states. Crystal Barrel, OBELIX and ASTERIX at LEAR provided the data on which the $f_0(1500)$ was established.

\paragraph{Heavy-flavour decays.}

Decays of $B$ and $D$ mesons at BaBar, Belle, Belle~II and LHCb produce light scalar and pseudoscalar systems with high statistics over wide kinematic ranges. They offer independent handles on the flavour content of the produced resonances through the flavour structure of the weak vertex.

\paragraph{Two-photon fusion as a veto.} 

As gluons carry no electric charge, a pure glueball does not couple to two photons. In $e^+e^-\to e^+e^-\gamma^*\gamma^*\to e^+e^- X$ a pure glueball should therefore be absent while ordinary $q\bar{q}$ mesons are produced. Combining this with the radiative $J/\psi$ rate gives a discriminating ratio. Two related quantities appear in the literature and are often confused. Stickiness \cite{chan84} is defined as
\begin{equation}
    S_R \;\propto\; \frac{\Gamma(J/\psi \to \gamma R)}{\Gamma(R\to\gamma\gamma)}\times \big(\text{phase-space and multipolarity factors}\big),
\end{equation}
with an arbitrary overall normalisation, so only ratios of stickiness between states are meaningful. Gluiness \cite{clos97} is the closely related quantity 
\begin{equation}
    G_R\; \propto\; \frac{\Gamma(R\to gg)}{\Gamma(R\to\gamma\gamma)},
\end{equation}
normalised so that $G_R = 1$ for an ordinary $q\bar{q}$ meson. A pure glueball should have $G_R \gg 1$. Measurements come from CLEO~II, ALEPH and L3 at LEP, and from Belle. This veto is powerful but not absolute: interference between a glueball and its $q\bar{q}$ mixing partners can suppress or enhance the $\gamma\gamma$ coupling of a physical state independently of its glueball fraction.

\section{Experimental status}
\label{sec:experiment}

In this final section preceding the concluding remarks, the experimental status of the main glueball candidates is discussed. The scalar, pseudoscalar, and tensor sectors are reviewed separately, while the limited experimental evidence for the observation of Oddballs through the Odderon is considered last.

\subsection{The scalar sector}
\label{sec:scalar}

The quark model admits exactly two isoscalar states per nonet: $\ket{n\bar{n}}$ and $\ket{s\bar{s}}$. \Cref{tab:f0} lists the established $I^G(J^{PC}) = 0^+(0^{++})$ resonances below $2.1$~GeV. There are six, and, after removing $f_0(500)$ and $f_0(980)$ which belong to the light scalar nonet and are widely understood as conventional quark-antiquark states~\cite{pela25}, one is left with $f_0(1370)$, $f_0(1500)$, $f_0(1710)$ and $f_0(2020)$ where the quark model wants two. That surplus is the original and still the strongest empirical motivation for a scalar glueball in this region, and it coincides with the mass range lattice QCD predicts.

\begin{table}[t]
    \centering
    \caption{Established isoscalar scalar resonances below $2.1$~GeV, from the Review of Particle Physics \cite{pdg24}. Both the $T$-matrix pole position and the Breit--Wigner parameters are given. For the broad states, the pole is the meaningful quantity and the Breit--Wigner values are strongly parameterisation-dependent. The glueball is generally taken to mix with $f_0(1370)$, $f_0(1500)$ and $f_0(1710)$, and in some analyses also with $f_0(2020)$. \justifying}
    \label{tab:f0}
    \vspace{2mm}
    \small
    \begin{tabular}{lrrr}
        \hline\hline
        Resonance\hspace{1cm} & \hspace{0.5cm}$T$-matrix pole (MeV) & \hspace{0.5cm} Breit--Wigner $M$, &$\Gamma$ (MeV) \\
        \hline
        $f_0(500)$   & $(400\text{--}550) - i(200\text{--}350)$ & $400\text{--}800$, & $100\text{--}800$ \\
        $f_0(980)$   & $(980\text{--}1010) - i(20\text{--}35)$  & $990\pm 20$, & $10\text{--}100$ \\
        $f_0(1370)$  & $(1250\text{--}1440) - i(60\text{--}300)$ & $1200\text{--}1500$, & $200\text{--}500$ \\
        $f_0(1500)$  & $(1430\text{--}1530) - i(40\text{--}90)$  & $1522\pm 25$, & $108\pm 33$ \\
        $f_0(1710)$  & $(1680\text{--}1820) - i(50\text{--}180)$ & $1733^{+8}_{-7}$, & $150^{+12}_{-10}$ \\
        $f_0(2020)$  & $(1870\text{--}2080) - i(120\text{--}240)$ & $1982^{+54}_{-3}$, & $440\pm 50$ \\
        \hline\hline
    \end{tabular}
\end{table}

This counting argument is, however, weaker in practice than in textbooks. No single reaction displays all the candidate states cleanly and simultaneously. They appear across different processes, with $f_0(1370)$ and $f_0(1500)$ prominent in $\bar{p}p$ annihilation, while $f_0(1500)$ and $f_0(1710)$ are favoured in radiative $J/\psi$ decay and central production. The status of $f_0(1370)$ is itself contested: it is broad and overlaps strongly with $f_0(1500)$. Several analyses find that it does not require a distinct pole. Extracting four overlapping poles from coupled channels is exactly the regime in which parameterisation bias dominates. Rather than a settled assignment, the field currently supports three positions.
\begin{itemize}

    \item The glueball is mainly $f_0(1500)$. This was the original interpretation, from Amsler and Close \cite{amsl95,clos96} and Close and Kirk \cite{clos01}. Its supporting evidence is that $f_0(1500)$ is comparatively narrow, is produced strongly in $\bar{p}p$ annihilation and in central production but is suppressed in $\gamma\gamma$ fusion. Its decay pattern into $\pi\pi$, $K\bar{K}$ and $\eta\eta$ is closer to flavour-blind than that of its neighbours. In this picture $f_0(1710)$ is predominantly $s\bar{s}$. The scenario is supported by some extended linear sigma model fits \cite{giac05}. Its principal difficulty is that it requires $M_G \simeq 1.5$~GeV, at the low end of Eq.~\eqref{eq:honest}.
    
    \item The glueball is mainly $f_0(1710)$. This is currently the most widely held view. The quenched lattice mass sits closer to $1.7$~GeV under the most common scale settings and mass-matrix fits with $M_G \simeq 1.7$~GeV assign to $f_0(1710)$ a glueball fraction above $60\%$ \cite{chen06b,jano14}. Chiral suppression mechanism described in \Cref{sec:flavour} also predicts a scalar glueball that prefers $K\bar{K}$ over $\pi\pi$, which is the observed pattern for $f_0(1710)$. This state is produced copiously in radiative $J/\psi$ decay, as expected for a glueball candidate. But the most quantitative support comes from the JPAC coupled-channel analysis of BESIII $J/\psi\to\gamma\pi^0\pi^0$ and $\gamma K^0_S K^0_S$ data \cite{roda22}. It determined seven scalar and tensor poles using a large ensemble of parameterisations to control model bias, and found that the hierarchy of couplings favours $f_0(1710)$ as the state with the largest glueball component. Additional support comes from BESIII's upper limit \cite{besi22a}
    \begin{equation}
    \frac{\mathcal{B}\big(f_0(1710)\to\eta\eta'\big)}{\mathcal{B}\big(f_0(1710)\to\pi\pi\big)} < 1.61\times 10^{-3} \quad (90\%\ \text{CL}),
    \end{equation}
    which is consistent with the Witten--Sakai--Sugimoto prediction $\Gamma(\eta\eta')/\Gamma(\pi\pi)\lesssim 0.04$ for a nearly pure glueball \cite{brun15}. As the measured bound lies more than an order of magnitude below the prediction, it should be regarded as a consistency check that the glueball interpretation survives, rather than a quantitative confirmation of it.

    \item The glueball is not any one state. Sarantsev, Denisenko, Thoma and Klempt \cite{sara21} analysed the same radiative $J/\psi$ data within a coupled-channel framework including a larger set of $f_0$ states and reached a different conclusion. According to their analysis, the glueball is not concentrated in a single resonance but is distributed over the whole scalar isoscalar spectrum from $f_0(1370)$ to $f_0(2100)$. The summed glueball fraction reaches $(78\pm 18)\%$, with a centroid at $M = 1865\pm 25$~MeV and a large width of $\Gamma = 370\pm 50$~MeV. In this picture, attempting to identify a single $f_0$ state as the glueball is an ill-posed task. Rather the appropriate interpretation is that the glueball has been observed as a broad component spread across several scalar-isoscalar states, with a mass consistent with lattice predictions.
\end{itemize}
That two careful coupled-channel analyses of the same BESIII data reach such different conclusions may itself be the most important fact about the scalar sector. The difference traces to the treatment of the broad background, of the number of poles admitted, and of the constraints imposed by other reactions. Until it is resolved, statements that the scalar glueball has been identified are premature.

\subsection{The pseudoscalar sector}
\label{sec:pseudoscalar}

The ground states in the pseudoscalar sector are the $\eta$ and $\eta'$. As the lattice glueball sits above $2.4$~GeV, the mixing of a $0^{-+}$ glueball with these two states should be suppressed by the large mass gap. The $\eta'$ mass is already explained by the $U(1)_A$ anomaly through the Witten-Veneziano mechanism, without invoking a glueball admixture. The lattice determination of Jiang \emph{et al.} \cite{jian23}, of the $\eta-\eta'$ mixing angle $|\theta| = 3.46(46)^\circ$, is consistent with a gluonic probability below $1\%$ in the $\eta$. Phenomenological $\eta$-$\eta'$-$G$ mixing analyses \cite{math10,math10b} likewise bound the gluonic content of the $\eta'$ at the level of a few percent, with the extracted value depending on the mixing scheme adopted. The glueball should therefore be sought in the higher-energy spectrum.

Three pseudoscalar isoscalars have been reported between $1.2$ and $1.5$~GeV: $\eta(1295)$, $\eta(1405)$ and $\eta(1475)$. The quark model expects only two radial excitations in this energy range. For two decades $\eta(1405)$ was accordingly a glueball candidate, supported by its non-observation in two-photon fusion at L3 and CLEO. Two objections have since accumulated. First, a glueball at $1.4$~GeV is roughly $1$~GeV below every lattice prediction for the $0^{-+}$ ground state. This is a much larger discrepancy than the scale-setting ambiguity can absorb. Second, Wu, Liu, Zhao and Zou \cite{wu12} showed that a $K^*\bar{K}K$ triangle singularity produces a narrow enhancement between the charged and neutral $K\bar{K}$ thresholds, which can make a single resonance appear at two different masses in different decay channels. This would simultaneously explain the anomalously large isospin violation observed in the decay of $\eta(1405/1475)$ to $3\pi$. The current majority view is that $\eta(1405)$ and $\eta(1475)$ are one state, and that the pseudoscalar glueball is not to be found here.

Attention has therefore shifted decisively above $2$~GeV. Using $(10087\pm 44)\times 10^6$ $J/\psi$ events, BESIII performed a partial-wave analysis of $J/\psi\to\gamma K^0_S K^0_S \eta'$ and determined the spin-parity of the $X(2370)$ for the first time \cite{besi24}. Its current measured properties are
\begin{equation}
\begin{aligned}
    J^{PC} &= 0^{-+}, \\
    M &= 2395 \pm 11\,(\text{stat})\,^{+26}_{-94}\,(\text{syst})\ \text{MeV}, \\
    \Gamma &= 188\,^{+18}_{-17}\,(\text{stat})\,^{+124}_{-33}\,(\text{syst})\ \text{MeV},
\end{aligned}
\end{equation}
with a statistical significance above $11.7\sigma$. The state had been seen earlier in $J/\psi\to\gamma\pi^+\pi^-\eta'$ \cite{besi11} and in $J/\psi\to\gamma K\bar{K}\eta'$ \cite{besi20}. The same programme has produced the $X(2600)$ \cite{besi22b} and evidence for further structures. The case for $X(2370)$ as the pseudoscalar glueball rests on four main pieces of evidence. It has the right quantum numbers ; It is produced in a gluon-rich channel ; It decays to $\eta'$-rich final states, as expected for a state coupling to the gluonic anomaly ; And its mass is in the right region.

The fourth leg is weaker than it is often made to sound. The mainstream quenched lattice value for the $0^{-+}$ ground state is $2.56\text{--}2.60$~GeV (see \Cref{tab:spectrum}), roughly $200$~MeV above the measurement. The honest statement is that $X(2370)$ is compatible with the pseudoscalar glueball and is the best candidate available up to now. But the mass agreement does not by itself discriminate, and the measured production rate should be compared with the lattice prediction $\mathcal{B}(J/\psi\to\gamma G_{0^{-+}}) = 2.31(90)\times 10^{-4}$ \cite{gui19} before the identification is regarded as established.

A related result deserves mention because it illuminates how the gluonic sector is disentangled. In the same $J/\psi\to\gamma\eta\eta'$ analysis that produced the $f_0(1710)\to\eta\eta'$ bound, BESIII observed an isoscalar state named $\eta_1(1855)$ with exotic $J^{PC} = 1^{-+}$, $M = 1855\pm 9$~MeV and $\Gamma = 188\pm 18$~MeV \cite{besi22a}. Since $1^{-+}$ has $C=+1$, it is not an oddball, and since it is exotic for $q\bar{q}$, it is naturally interpreted as a hybrid meson \cite{dude26} rather than a glueball due to its mass lying in the range of the lattice results for the isoscalar hybrid meson. This first isoscalar exotic seen in a gluon-rich channel demonstrates that the experimental techniques required for the glueball programme are now in place.

\subsection{The tensor sector}
\label{sec:tensor}

The tensor glueball should be the most copiously produced of all in radiative $J/\psi$ decay, with $\mathcal{B}\approx 1.1\%$ predicted on the lattice \cite{yang13} (see \cref{sec:radiative}). It should also be the state that anchors the Pomeron trajectory: extrapolating $\alpha_P(t) = 1.08 + 0.25\,t$ to the first physical point, $J = 2$ at $t = M^2$, gives $M \approx 2.1\text{--}2.2$~GeV, close to but somewhat below the lattice value which is about $2.4$~GeV.

Against these strong expectations, the experimental situation is unresolved, for a straightforward reason: above $2$~GeV the tensor spectrum is dense. Radial and orbital $q\bar{q}$ excitations, tensor hybrids and possible multiquark states all crowd the same region, the widths exceeding the spacings. The historical candidate $f_J(2220)$ - also denoted $\xi(2230)$, reported by MARK~III and by BES with a narrow width and with two-photon suppression, was never confirmed at the statistics later available and is no longer listed as established. More recent analyses of BESIII radiative $J/\psi\to\gamma\pi^0\pi^0$ and $\gamma K^0_S K^0_S$ data by Klempt \emph{et al.} \cite{klem22} find a tensor enhancement described by a pole at $(2210\pm 60) - i(180\pm 60)$~MeV. In parallel with their scalar-sector conclusion, they suggest that the tensor glueball may be distributed among several high-mass tensor mesons rather than identified with one. The alternative that $f_2(1950)$ carries the dominant fraction has also been advanced \cite{klem23}. The JPAC analysis \cite{roda22} extracted tensor poles from the same data with a different parameterisation strategy. In short, the tensor sector is the clearest example of the general problem: the theoretical prediction is sharp, the production channel is right, the data exist, but the extraction of poles from a dense overlapping spectrum is the bottleneck.

\subsection{Oddballs and the Odderon}
\label{sec:oddballs}

The $C=-1$ sector is essentially unexplored. From \Cref{tab:spectrum}, the lightest $C=-1$ glueball in quenched lattice QCD is the $1^{+-}$ at $2.942(41)$~GeV, followed by states near $3.5$~GeV. The state that would sit on the Odderon trajectory, namely the $1^{--}$, is much heavier, around $3.8\text{--}4.0$~GeV, and the exotic $0^{+-}$ is heavier still, near $4.5$~GeV. Constituent three-gluon calculations in the helicity formalism \cite{chev25} reproduce the broad features of this pattern while predicting some additional low-lying states.

The connection between oddballs and high-energy scattering is the Odderon, the $C=-1$ partner of the Pomeron. It corresponds to the exchange of a $C$-odd colour-singlet, which in the constituent picture is a three-gluon exchange and which in Regge language is the leading $C=-1$ glueball trajectory. Its existence follows from QCD, but its intercept is the question. Llanes-Estrada, Bicudo and Cotanch \cite{llan06} computed the $J^{--}$ glueball trajectory in a Coulomb-gauge many-body approach and found an intercept well below unity, consistent with the smallness of the observed $C$-odd effects. Experimentally, the comparison of $pp$ and $p\bar{p}$ elastic scattering at the LHC and the Tevatron by TOTEM and D0 has been interpreted as evidence for Odderon exchange \cite{tote21}. If confirmed, it would be the only experimental handle on the $C=-1$ gluonic sector to date. Of course, this result does not identify any resonance directly, but constrains the $C=-1$ glueball trajectory in the scattering domain where $t\le 0$.

Direct searches for oddball resonances face compounding difficulties. Because the masses are high, production requires substantial energy. The decays are to high-multiplicity final states, such as $4\pi$, $6\pi$, $K\bar{K}\pi\pi$, and $J/\psi\,\pi\pi$ for the heaviest states, with severe combinatorial backgrounds from identical particles. The partial-wave analyses required involve many overlapping isobar channels ($\rho\rho$, $f_0\rho$, $a_1\pi$, $K^*\bar{K}^*$), where broad resonances and non-resonant continuum contributions introduce large model dependence.

As done for the positive charge conjugation sector in \cref{sec:selection}, the accessible constructions are worth stating explicitly, since they determine where to look. For a neutral pair of pseudoscalars, $C = (-1)^L$. Therefore a $C=-1$ pair requires odd relative orbital angular momentum. In practice, one instead builds $C=-1$ systems from an odd number of $C=-1$ isobars, such as a $\rho$ or $\omega$ recoiling against a $C=+1$ system. Typical examples are $\rho\,f_0$ or $\omega\,f_0$ in $4\pi$, and $\rho\,\eta$ or $\omega\,\eta$ in $\pi\pi\eta$. Radiative $J/\psi$ decay is not available here, since it produces $C=+1$ intermediary states. The accessible channels are direct $e^+e^-$ annihilation for $1^{--}$, hadronic $J/\psi$ and $\psi(2S)$ decays, central production, and $\bar{p}p$ annihilation in flight at PANDA.

\section{Outlook}
\label{sec:outlook}

Fifty years after they were conjectured, glueballs occupy an uncomfortable position: theoretically compulsory, computationally well characterised, and experimentally unconfirmed. What is solid is the pure-gauge spectrum. Independent methods that are lattice simulation, functional equations, constituent models in the helicity basis, Coulomb-gauge many-body theory, holographic duals and sum rules agree on the ordering, on the ratio $M_{2^{++}}/M_{0^{++}} \approx 1.4$, and on the gross scale, with the lattice supplying dimensionless ratios at the percent level. The confrontation with experiment is limited by two structural facts: that converting a quenched mass to GeV is ambiguous at the $10\text{--}15\%$ level (see \Cref{sec:scale}), and that the physical states are broad, overlapping resonances rather than the stable states the calculations describe. Progress therefore requires work on three fronts.

On the lattice, the decisive step is the treatment of glueballs as resonances: finite-volume coupled-channel calculations in the scalar isoscalar sector, with $\pi\pi$, $K\bar{K}$, $\eta\eta$ and $4\pi$ included, using distillation and the variational method with both gluonic and $q\bar{q}$ operators. This is technically formidable but no longer out of reach, and it would replace the mass-matrix phenomenology of \Cref{sec:mixing} with a first-principles calculation. Continued work on radiative production matrix elements in the unquenched theory is equally valuable, since those rates are the sharpest experimental targets.

On the amplitude-analysis side, the disagreement between the JPAC and Bonn--Gatchina analyses of the same BESIII data (\Cref{sec:scalar}) is the most urgent problem in the field, and it is a problem about method rather than about statistics. Resolving it requires analyses that impose analyticity and unitarity systematically, that quantify parameterisation bias explicitly, and that fit several reactions simultaneously rather than one at a time.

On the experimental side, BESIII's $\sim 10^{10}$ $J/\psi$ sample is not yet fully exploited, and a future super tau-charm facility would extend it by another order of magnitude while adding polarisation. LHCb and the LHC forward programmes bring central exclusive production to energies where double-Pomeron dominance is cleaner than at the SPS. Finally, GlueX and CLAS12 at Jefferson Lab (and in the longer term the Electron--Ion Collider) address the neighbouring hybrid sector, whose resolution will sharpen the glueball problem by removing one class of competing interpretation.

The most likely outcome is not a single decisive discovery but a gradual quantitative convergence. This would emerge from consistent glueball fractions extracted from several reactions by mutually consistent analyses, and matched against unquenched lattice calculations of the same quantities. On present evidence, the scalar glueball is somewhere in the $f_0(1370)$-$f_0(2020)$ complex, quite possibly spread across it, and $X(2370)$ is a strong pseudoscalar candidate. Turning either statement into a measurement is the work ahead. None of the above three fronts can deliver the answer alone, and this is the central practical lesson of fifty years of glueball physics. A lattice mass without an amplitude analysis to compare it to is a number without a referent. An amplitude analysis without a theoretical prediction for couplings and production rates determines poles but not structure. An experiment without both is a measurement in search of an interpretation. Progress has come, and will continue to come, from the deliberate coordination of the three, which requires defining the same quantities on each side, quantifying systematic uncertainties honestly, and comparing like with like.

It is worth closing with the key message from \Cref{sec:whatmatters}. The object of the exercise is not to award a title to one resonance. It is to understand the physical spectrum of hadrons and the couplings that govern how they interact. In other words, it is to be able to say why that spectrum looks the way it does. Glueballs are a means to those ends: a hypothesis that, if established quantitatively, would explain a set of otherwise puzzling features of the light meson spectrum and would connect them directly to the mass gap of Yang-Mills theory. That is what would be gained, and it is worth the effort it is taking.

\end{document}